\documentclass[letterpaper, 10 pt, conference]{ieeeconf}  

\IEEEoverridecommandlockouts                              

\usepackage{graphics} 
\usepackage{epsfig} 
\usepackage{mathptmx} 
\usepackage{times} 
\usepackage{amsmath} 
\usepackage{amssymb}  
\usepackage{amsfonts}

\usepackage{comment}
\usepackage{stmaryrd}
\usepackage{bm}

\usepackage{cite}
\usepackage{url}

\renewcommand{\baselinestretch}{0.991}

\makeatletter
\newcommand\subparagraph{%
  \@startsection{subparagraph}{5}
  {\parindent}
  {3.25ex \@plus 1ex \@minus .2ex}
  {-1em}
  {\normalfont\normalsize\bfseries}}
\makeatother
\usepackage{titlesec}
\let\subparagraph\relax 

\titlespacing*{\section}{0pt}{*0.9}{*0.8}
\titlespacing*{\subsection}{0pt}{*0.9}{*0}

\usepackage{mathtools}

\usepackage{accents}
\newcommand{\ubar}[1]{\underaccent{\bar}{#1}}
\newcommand{\RNum}[1]{\uppercase\expandafter{\romannumeral #1\relax}}

\usepackage[noend]{algpseudocode}
\usepackage{algorithm}

\usepackage{etoolbox}
\makeatletter
\patchcmd{\@makecaption}
  {\\}
  {.\ }
  {}
  {}
\makeatletter
\patchcmd{\@makecaption}
  {\scshape}
  {}
  {}
  {}
\makeatother

\title{
LFT Representation of a Class of Nonlinear Systems:\\ A Data-Driven Approach}

\author{Sourav Sinha, Devaprakash Muniraj, and Mazen Farhood
\thanks{The authors are with the Kevin T. Crofton Department of Aerospace and Ocean Engineering, Virginia Tech, Blacksburg, VA 24061, USA. Email: \{srvsinha, devapm, farhood\}@vt.edu. This work was supported by the National Science Foundation (NSF) under Grant No. CMMI-1351640, the Office of Naval Research under Award No. N00014-18-1-2627, and the Center for Unmanned Aircraft Systems (C-UAS), an NSF Industry/University Cooperative Research Center (I/UCRC) under NSF~Grant~No.~CNS-1650465.}}

\begin{document}

\maketitle
\thispagestyle{empty}
\pagestyle{empty}

\begin{abstract}
This paper focuses on developing a method to obtain an uncertain linear fractional transformation (LFT) system that adequately captures the dynamics of a nonlinear time-invariant system over some desired envelope.
%
%
%
%
First, the nonlinear system is approximated as a polynomial nonlinear state-space (PNLSS) system, and a linear parameter-varying (LPV) representation of the PNLSS model is obtained. To reduce the potentially large number of scheduling parameters in the resulting LPV system, an approach based on the cascade feedforward neural network (CFNN) is proposed. We account for the approximation and reduction errors  through the addition of a norm-bounded, causal, dynamic uncertainty in the  LFT system. Falsification is used in a novel way to perform guided simulations for deriving a norm bound on the dynamic uncertainty and generating data for training the CFNN. Then, robustness analysis using integral quadratic constraint theory is carried out to choose an LFT representation that leads to a computationally tractable analysis problem and useful analysis results. Finally, the proposed approach is used to obtain an LFT representation for the nonlinear equations of motion of a fixed-wing unmanned aircraft system.
\end{abstract}
\vspace{-0pt}
\section{INTRODUCTION} \label{introduction}

%
%
%
%
This paper deals with constructing uncertain linear systems formulated in a linear fractional transformation (LFT) framework \cite{Doyle1996} that approximately capture nonlinear system behavior over some desired envelope. Based on the type of uncertainties, 
methodologies such as $\mu$-analysis \cite{Doyle1996_2} and integral quadratic constraint (IQC) theory \cite{Megretski1997} can then be used to analyze the robust performance of these uncertain LFT systems. 
%
%
Herein, we consider nonlinear time-invariant state-space systems, where the equations governing the system dynamics and outputs are nonlinear in the state and input variables. The goal is to represent such a system over some desired envelope with an LFT system that is an interconnection of a nominal linear time-invariant (LTI) model and a perturbation operator consisting of static linear time-varying (SLTV) uncertainties and a norm-bounded, causal, dynamic uncertainty. The purpose behind obtaining this LFT representation is to be able to conduct IQC-based analysis. In addition to SLTV uncertainties, the IQC-based analysis approach can handle dynamic uncertainties that are LTI, linear time-varying, or possibly nonlinear \cite{veenman2016}.
The SLTV uncertainties are used to model the system nonlinearities, wherein the uncertain parameters consist of state and input variables. The LFT with the SLTV uncertainties is obtained by first approximating the nonlinear system with a polynomial linear parameter-varying (LPV) system
%
%
%
%
and then expressing the LPV system as an LFT. The dynamic uncertainty accounts for the error incurred by representing the nonlinear system as an LFT on the SLTV uncertainties. Note that robustness analysis will still be required even when dealing directly with the nonlinear system, as it is very likely that the nonlinear system will not be an exact representation of the physical process due to the deliberate or inevitable undermodeling.



Various methods to obtain a polynomial LPV  representation of a nonlinear system exist; see \cite{Toth2010}.
We adopt the function substitution method, which requires the nonlinear system to be expressed in polynomial form.
%
%
This is achieved by combining the methods described in \cite{Paduart2010} and \cite{Brunton2016} to approximate the nonlinear system with a polynomial nonlinear state-space (PNLSS) system that is valid within an envelope. As in \cite{Paduart2010}, we use monomials in state and input variables as basis functions to model the system nonlinearities.
%
%
%
%
To estimate the coefficients corresponding to each basis function, we solve regression problems using the input-output data as in \cite{Brunton2016}, albeit with a crucial difference.
%
%
%
%
The number of basis functions and consequently the size of the optimization variable increase factorially with the number of state and input variables. As a result, the preceding method becomes ill-suited for high-dimensional systems. We overcome this difficulty by leveraging the known structure of the nonlinear system to define distinct basis functions for modeling the nonlinearities in each of the state and output equations.
%
%
%
%

For high-dimensional systems,
%
%
%
%
this approach results in an LPV system with a large number of coupled parameters, thereby leading to an analysis problem that might not be computationally tractable and could result in overly conservative analysis results \cite{Werner2008}.
%
%
In general,
a low-dimensional encoding of high-dimensional datasets can be found, as many high-dimensional systems of interest evolve on or near a low-dimensional manifold that is embedded in the original high-dimensional space \cite{Maaten2009}.
%
%
In \cite{Koelewijn}, a comparison of methods based on principal component analysis (PCA), kernel PCA, autoencoders, and deep neural networks for reducing the number of parameters in an affine LPV system is presented. Since the polynomial LPV system obtained from the PNLSS model is more general than the affine LPV system considered in \cite{Koelewijn}, the preceding methods are not directly applicable.
%
%

%
%
%
%
We propose an approach based on the cascade feedforward neural network (CFNN) to reduce the number of parameters in the polynomial LPV system. The CFNN-based approach
%
%
enables one
to directly minimize the error between the states and outputs of the original and the reduced LPV systems while reducing the number of parameters.
This reduction process  inevitably leads to loss of information and introduces some error.
%
%
To account for this reduction error and the PNLSS model estimation error, we add a dynamic uncertainty to the LFT system with SLTV uncertainties. Different uncertain LFT systems can be obtained based on the number of SLTV parameters reduced using the CFNN framework.
We conduct a trade-off study using IQC analysis to choose an uncertain LFT system that strikes a balance between the computational complexity of the resulting analysis problem and the conservativeness of the analysis results.

Data generation is a key part of the proposed data-driven approach. One needs data to estimate the PNLSS model, to train the CFNN, and to estimate the norm bound on  the dynamic uncertainty. For PNLSS model estimation, we use the Halton method to sample data from the envelope, as it has been shown to generate more uniform samples in a high-dimensional sample space compared to other data sampling methods \cite{Perdegnana2016}. As for generating data for the other two steps, we make use of falsification \cite{breach, staliro},
%
%
which is typically used in the context of finding initial conditions and input signals resulting in system trajectories that violate a given property expressed as a temporal logic formula.
%
%
%
%
Herein, falsification is used in a novel way to conduct guided simulations, and compared to naive Monte Carlo simulations, it affords the following benefits: a) one is likely to obtain an improved estimate of the norm bound on the dynamic uncertainty; and b) one can generate system trajectories that uniformly cover the envelope for parameter reduction.
%
%

%
%
The rest of the paper is organized as follows: Section \RNum{2} gives the notation and some needed definitions; Sections \RNum{3} and \RNum{4} describe how to obtain the LPV representation of the nonlinear system and  solve the parameter reduction problem, respectively; Section \RNum{5} presents the method used to quantify the dynamic uncertainty and the trade-off study using IQC analysis; Section \RNum{6} presents a case study involving the nonlinear equations of motion of a fixed-wing unmanned aircraft system (UAS); Section \RNum{7} gives some  concluding~remarks.
\section{PRELIMINARIES} \label{preliminaries}

\subsection{Notation}
The sets of non-negative integers, real vectors of  $n$ elements, and real matrices of size $n\times m$ are denoted by $\mathbb{Z}_+$, $\mathbb{R}^n$, and $\mathbb{R}^{n \times m}$, respectively. The $m \times m$ identity matrix  is denoted by $I_m$. The set $\{1,\,2,\,\dots,\,p\}$ is denoted by $\mathbb{N}_p$.
%
%
The block diagonal augmentation of matrices $A_1,\, A_2,\, \dots,\, A_n$ is given by $\text{diag}(A_1,\, A_2,\, \dots,\, A_n)$.
Given a matrix $A \in \mathbb{R}^{n \times m}$, $\left\lVert A \right\Vert_{F}$ denotes its Frobenius norm. Given $u\in\mathbb{R}^n$, its $1$-norm and $2$-norm are defined as $\lVert u \rVert_{1} = \sum_{i=1}^{n}\lvert u_i \rvert$ and $\lVert u \rVert_{2} = \sqrt{u^Tu}$, respectively.
%
%
A hyperrectangle is denoted by the set $\mathbb{H} = \{(p_x, p_u)\in \mathbb{R}^{n} \times \mathbb{R}^{n_u}  \mid  \ubar{p}_{x,i} \leq p_{x,i} \leq \bar{p}_{x,i}, \ \ubar{p}_{u,j} \leq p_{u,j} \leq \bar{p}_{u,j} \ \text{for all}\ i = 1, \dots, n \ \text{and} \ j = 1, \dots, n_u\} $.
$\ell_2^n$ denotes the normed space of square summable vector-valued sequences $x=(x(0),\,x(1),\,\dots)$, with each $x(k)\in \mathbb{R}^n$. In the sequel, $\ell_2^n$ is abbreviated as $\ell_2$ when the dimension is irrelevant to the discussion. Given $x\in \ell_2$, its $\ell_2$-norm is defined as $\lVert x \rVert_{\ell_2}^2 = \sum_{k=0}^{\infty}x(k)^Tx(k) $.
For Hilbert spaces $\mathcal{W}$ and $\mathcal{V}$, the $\mathcal{W}$-to-$\mathcal{V}$-induced norm of a bounded linear operator $P$ mapping $\mathcal{W}$ to $\mathcal{V}$ is defined as $\lVert P \rVert_{\mathcal{W}\rightarrow \mathcal{V}} =      \sup_{0\neq w\in \mathcal{W}} ({\lVert Pw \rVert_{\mathcal{V}}}/{\lVert w \rVert_{\mathcal{W}}})$.
%
%

\subsection{LPV and LFT Systems}
%
%
A discrete-time LPV system is defined as
\begin{equation}
    \begin{bmatrix}\bar{x}(k+1) \\ \bar{y}(k) \end{bmatrix} =
    \begin{bmatrix}A(\rho(k)) & B(\rho(k)) \\ C(\rho(k)) & D(\rho(k)) \end{bmatrix}
    \begin{bmatrix}\bar{x}(k) \\ \bar{u}(k) \end{bmatrix}
    \label{eqn2_2_1}
\end{equation}
where $k \in \mathbb{Z}_+$, $\bar{x}(k) \in \mathbb{R}^n$ is the state, $\bar{y}(k) \in \mathbb{R}^{n_y}$ is the output, $\bar{u}(k) \in \mathbb{R}^{n_u}$ is the input, and $\rho(k)$ is the time-varying parameter (also known as the scheduling parameter).
The parameter $\rho(k)\coloneqq [\rho_1(k),\,\rho_2(k),\, \dots,\,\rho_l(k)]^T$ and the parameter increment $d\rho(k)  \coloneqq \rho(k{+}1)-\rho(k)$ are such that $(\rho(k),d\rho(k)) \in \Gamma$ for all $k\in \mathbb{Z}_{+}$, where $\Gamma$ is defined as $\Gamma \coloneqq \{(p,dp) \in \mathbb{R}^l\times \mathbb{R}^l \mid \ubar{p}_{i} \leq p_{i} \leq \bar{p}_{i}, \ \ubar{d}p_{i} \leq dp_{i} \leq \bar{d}p_{i} \ \text{for all}\ i = 1,\,2,\,\dots,\, l \}$.
%
%
A polynomial LPV system is one where the state-space matrix-valued functions have polynomial dependence on the scheduling parameters. The LPV system in \eqref{eqn2_2_1} can be equivalently represented in LFT form as (see Fig.~\ref{fig_rLFT_wDLTI}a)
\begin{equation}
    \begin{bmatrix}\bar{x}(k{+}1)^T &\!\!\!\! \varphi(k)^T &\!\!\!\! \bar{y}(k)^T \end{bmatrix}^T =
    G
    \begin{bmatrix}\bar{x}(k)^T &\!\!\!\! \vartheta(k)^T &\!\!\!\! \bar{u}(k)^T \end{bmatrix}^T
    \label{eqn2_2_2}
\end{equation}
where $\vartheta = \Delta \varphi$, $G$ is the nominal LTI system, and
$\Delta = \text{diag}(\rho_1 I_{r_1},\,\rho_2 I_{r_2},\,\dots,\,\rho_l I_{r_l})$
is the perturbation operator.

\section{LPV REPRESENTATION}\label{discrepancy_modeling}

%
%
The nonlinear state-space systems considered in this work are of the following form:
\begin{subequations}
\label{eqn3_0}
\begin{align}
\dot{x}(t) &= f(x(t),u(t)),~~~x(t_0)=x_0 \label{eqn3_0_1} \\
y(t) & = h(x(t),u(t)) \label{eqn3_0_2}
\end{align}
\end{subequations}
where $x(t) \in \mathbb{R}^n$, $y(t) \in \mathbb{R}^{n_y}$, and $u(t) \in \mathbb{R}^{n_u}$ are, respectively, the state, output, and input vectors of the system.
%
%
%
%
Linearizing equation \eqref{eqn3_0_1} about the operating point $(x^{*}, u^{*})$ and then discretizing it using zero-order hold sampling with a sampling time of $\tau\,\, \mathrm{s}$ yield the  discrete-time LTI state equation
\begin{equation} \bar{x}(k+1) \approx A_d\bar{x}(k)+B_d\bar{u}({k}) \label{eqn3_1_2} \end{equation}
where $k \in \mathbb{Z}_+$, $\bar{x}(k)=x(k\tau)-x^{*}$, and $\bar{u}(k)=u(k\tau)-u^{*}$.

\subsection{PNLSS Model Estimation}

To obtain the PNLSS model, we use polynomial functions to model the nonlinearities ignored during linearization.
%
%
We choose monomials in $\bar{x}$ and $\bar{u}$ of degree varying from $2$ to $p$ as the basis functions, which are defined as $\prod_{i=1}^{n+n_u} \bar{z}_i^{\alpha_{ji}} \text{ for } j \in \mathbb{N}_{n_{p}}$,
where $\bar{z}_i$ is the $i^{th}$ element of $\bar{z}$ with $\bar{z}^T = [\bar{x}^T \ \bar{u}^T]$, ${n_{p}}$ is the total number of basis functions, and for $j \in \mathbb{N}_{n_{p}} $, $2 \leq \sum_{i=1}^{n+n_u}  \alpha_{ji} \leq p$.
%
%
As noted earlier, the preceding approach is not suitable for high-dimensional systems since the number of basis functions increases factorially with the size of $\bar{z}$.
For  physical systems of interest, the nonlinearities in the system depend only on a few variables, and two equations may or may not depend on the same variables.
Leveraging the fact that the structure of the nonlinear function $f$ is known, we can define distinct nonlinear basis functions for each of the state equations, thereby reducing the computational complexity of the regression problem.
%
%
The number of basis functions can be further reduced by placing additional constraints on each variable in the monomials.
Using a vector-valued nonlinear function $\zeta_k\,{:}\,  \mathbb{R}^n\times  \mathbb{R}^{n_u}\rightarrow  \mathbb{R}^{n_{p_k}}$ that maps $\bar{x}$ and $\bar{u}$ to the monomials chosen to model the nonlinearities in the $k^{th}$ state equation, where ${n_{p_k}}$ is the number of monomials for the $k^{th}$ state, we model the
%
%
ignored nonlinearity as
$ E^T\zeta (\bar{x},\bar{u})$, where ${\zeta}(\bar{x},\bar{u})=[{\zeta}_1(\bar{x},\bar{u})^T \ {\zeta}_2(\bar{x},\bar{u})^T \ \dots \ {\zeta}_n(\bar{x},\bar{u})^T]^T$
and ${E} = \text{diag}({E}_1,\, {E}_2,\, \dots,\, {E}_n) \in \mathbb{R}^{{n_p} \times n} $ with ${E}_k \in R^{n_{p_k}}$ being the coefficient vector corresponding to the $k^{th}$ state and $n_p = \sum_{k=1}^{n} n_{p_k}$. The arguments of $\bar{x}$ and $\bar{u}$ are suppressed for convenience.

%
%
To determine the coefficient matrix $E$ from data, we sample $(p_x(i),p_u(i))$ for $i\in \mathbb{N}_s$ from the user-defined envelope $\mathbb{H}$ within which one intends to approximate the nonlinear system, and define $y_k$ and $\Theta_k$ for $k\in\mathbb{N}_n$ as
\begin{align*}
y_k &= [\delta{x}_k(1) \,\, \delta{x}_k(2) \,\,  \dots \,\, \delta{x}_k(s)]^T \ \text{and} \\
\Theta_k &= [\zeta_k\!\left(p_x(1),p_u(1)\right) \,\, \zeta_k\!\left(p_x(2),p_u(2)\right) \, \dots \, \zeta_k\!\left(p_x(s),p_u(s)\right)]^T
\end{align*}
where $\delta x(i)$ is the discrepancy between $x(\tau)$ obtained from the nonlinear system \eqref{eqn3_0_1} and the linear system \eqref{eqn3_1_2}, given $\bar{x}(0) = p_x(i)$, $\bar{u}(0) = p_u(i)$, and $u(t)$ is constant for all $t\in[0,\tau)$.
As in \cite{Brunton2016}, we formulate the regression problem as $n$ unconstrained optimization problems given by
\begin{equation}
\begin{aligned}
\min_{E_k} \quad & \left\lVert y_k - \Theta_k E_k \right\Vert_{2} + \sigma\left\lVert E_k \right\Vert_{1} \ \text{for}\ k \in \mathbb{N}_{n}.
\end{aligned}
\label{eqn3_3_1}
\end{equation}
However, unlike \cite{Brunton2016}, the nonlinear basis functions $\Theta_k$ in our case are different for each $k$ since distinct basis functions are used for each state equation.
$\sigma \lVert E_k   \rVert_1$ is a regularization term used to search for a sparse solution. We can now say that for all $(\bar{x}({k}), \bar{u}({k})) \in \mathbb{H}$, the following PNLSS system
\begin{equation} \bar{x}(k+1) = A_d\bar{x}(k)+B_d\bar{u}({k}) + {E}^T{\zeta}\left(\bar{x}({k}),\bar{u}({k})\right)
\label{eqn3_3_3}
\end{equation}
is a close approximation of the nonlinear system \eqref{eqn3_0_1}.
\begin{figure}[tb]
\centering
\includegraphics[width=0.31\textwidth]{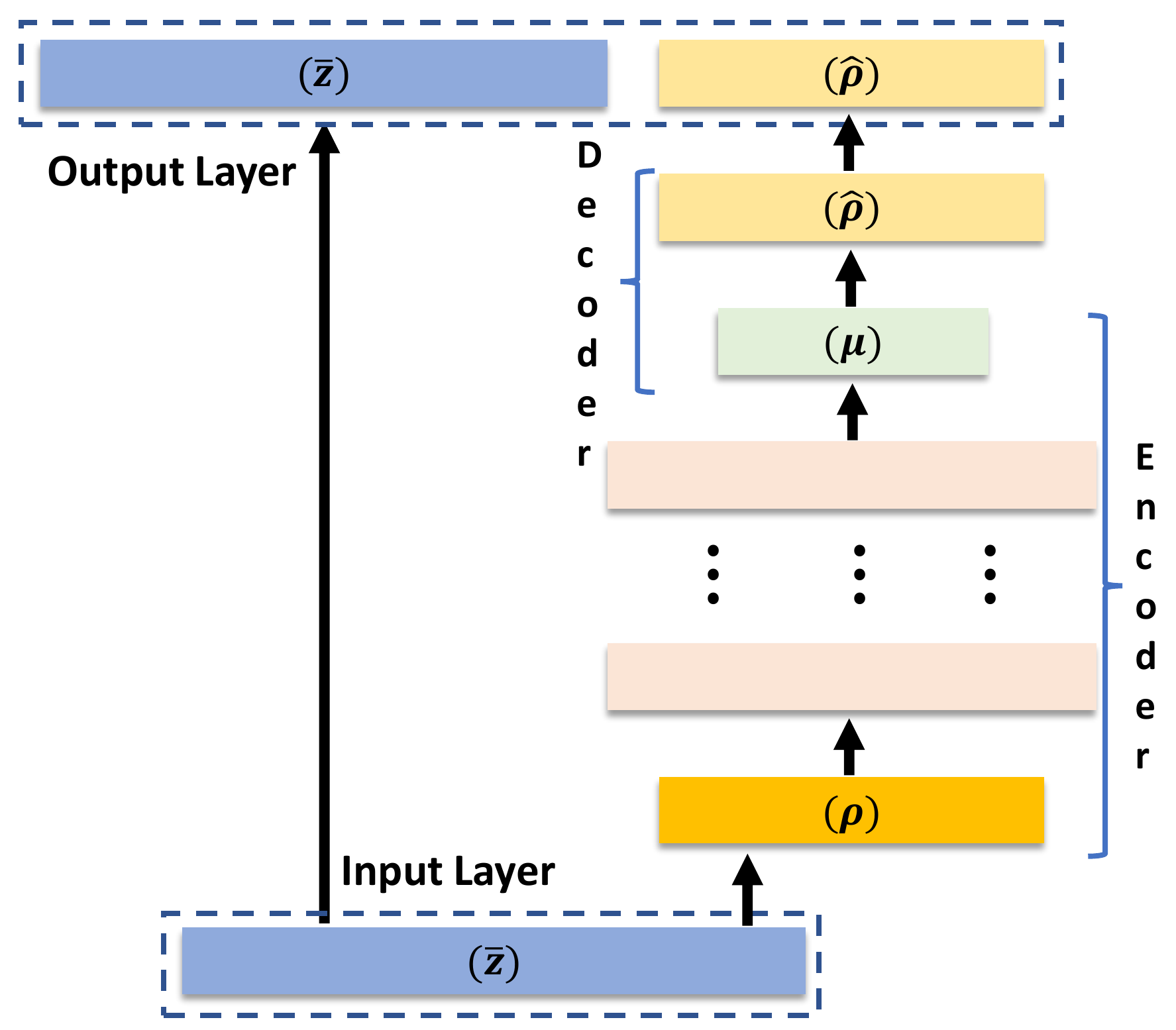}
\caption{CFNN-based parameter reduction framework.}
\label{fig_CFNNframework}
\end{figure}

\subsection{LPV Representation of the PNLSS Model} \label{LPV_embedding}

The following LPV model is obtained from \eqref{eqn3_3_3} through an exact transformation that does not induce any errors:
\begin{equation}
{\bar{x}}(k+1) = A (\rho(k))\bar{x}(k)+B (\rho(k))\bar{u}(k)
\label{eqn4_0_1}
\end{equation}
where
%
%
%
%
$A (\rho) = A_d + \Delta A_d(\rho)$ and $B (\rho) = B_d + \Delta B_d(\rho)$, and $\rho(k) \in \mathbb{R}^l$ is the vector of scheduling parameters (evaluated at discrete instant $k$) consisting of a subset of the state and input variables. The
%
%
%
%
matrices $\Delta A_d(\rho)$ and $\Delta B_d(\rho)$ are obtained by decomposing $ {E}^T{\zeta}(\bar{x},\bar{u})$ as $\Delta A_d(\bar{x},\bar{u})\bar{x} + \Delta B_d(\bar{x},\bar{u})\bar{u}$.
%
%
%
%
The number of different ways in which each of the monomial bases can be factorized is equal to the number of state and input variables present in the term. To automate the process, we define a priority order to factor out variables from the monomials.
%
%
The priority order is generated with the goal of minimizing the number of parameters in the LPV system. Once the priority order is finalized, we factorize ${\zeta}(\bar{x},\bar{u})$ as $Q\bar{z}$ and $[\Delta A_d(\rho) \quad \Delta B_d(\rho)] \equiv {E}^TQ$.
Defining $ind_k$ to be the index of the element of $\bar{z}$ that is factored out from the $k^{th}$ element of ${\zeta}(\bar{x},\bar{u})$, the elements of $Q$ can be written~as
\[
  Q_{kj} =
  \begin{cases}
     \left(\displaystyle\prod_{i=1}^{n+n_u} \bar{z}_i^{\alpha_{ki}}\right)\bar{z}_{ind_k}^{-1} & \text{if $j=ind_k$} \\
      \hspace{10mm} 0 & \text{if $j\neq ind_k$}.
  \end{cases}
\]
%
%
Following the method discussed in this section, one can also obtain an LPV representation
 of the output equation \eqref{eqn3_0_2}.
%
%

\section{REDUCING THE NUMBER OF PARAMETERS}
\label{SDR}

%
%
To reduce the potentially large number of tightly-coupled parameters in the resulting LPV system, we employ a learning-based parameter reduction technique.
%
%
Given an LPV system as in \eqref{eqn2_2_1} with $l$ parameters, our goal is to find a mapping $\eta: \mathbb{R}^l\rightarrow \mathbb{R}^m$ with $m<l$ and an inverse mapping $\eta^\ast: \mathbb{R}^m\rightarrow \mathbb{R}^l$ such that the following LPV system
%
%
\begin{subequations}
\label{eqn5_1}
\begin{align}
{\bar{x}}(k+1)  & =\hat{A} (\mu(k))\,\bar{x}(k)+\hat{B}  (\mu(k))\,\bar{u}(k) \label{eqn5_1_3}\\
{\bar{y}}(k)  & =\hat{C} (\mu(k))\,\bar{x}(k) + \hat{D} (\mu(k))\,\bar{u}(k) \label{eqn5_1_4}
\end{align} \end{subequations}
with $\mu  =  \eta(\rho)$ satisfactorily approximates the original $l$-parameter LPV system. The time dependence of $\rho$ and $\mu$ has been suppressed for convenience. The system matrices in \eqref{eqn5_1} are obtained using the following relation:
\begin{equation}
    \hat{M}(\mu) = \begin{bmatrix}
    \hat{A} (\mu) & \hat{B} (\mu)\\
    \hat{C} (\mu) & \hat{D} (\mu)
    \end{bmatrix}  =
    \begin{bmatrix}
    {A} (\hat{\rho}) & {B} (\hat{\rho})\\
    {C} (\hat{\rho}) & {D} (\hat{\rho})
    \end{bmatrix} = M(\hat{\rho}).
    \label{eqn5_1_6}
\end{equation}
$\hat{\rho}$ is an estimate of the original parameter $\rho$ and is reconstructed from $\mu$ as $\hat{\rho} = \eta^\ast(\mu)$. Since $\hat{\rho}$ is an approximation of $\rho$, we have $\hat{M}(\mu) = M(\hat{\rho}) \approx {M}(\rho)$.
%
%
To obtain the mappings $\eta$ and $\eta^{*}$, we propose a CFNN-based framework; see Fig.~\ref{fig_CFNNframework}. The framework resembles an artificial neural network, albeit with the following difference: the output layer, in addition to being connected to the last hidden layer, is also connected directly to the input layer. The data set $\Xi \in \mathbb{R}^{(n+n_u) \times N}$, which is defined as 
$\begin{bmatrix} w(1) &\!\!\! w(2) &\!\!\! \dots &\!\!\! w(N)\end{bmatrix} $,
where $w(i)$ for $i\in\mathbb{N}_N$ are samples of $\bar{z}$, is constructed by stacking together the state and input histories from multiple trajectories of the nonlinear system and is fed as the input to the network.
%
%
%
%
To effectively learn the lower-dimensional encoding of the scheduling parameters, it is important to generate system trajectories that uniformly cover the expected range of operation. These trajectories can either be generated through experiments or simulations. In Section~\ref{casestudy}, we describe a method that employs falsification, which is briefly discussed in Section~\ref{reductionerror}, to efficiently generate such trajectories through guided simulations.

\vspace{-0mm}
The first part of the network is the encoder that encodes the original higher-dimensional data in a lower-dimensional space. Multiple layers can be used in the encoder to capture complex nonlinearities in the dataset. We use $n_e$ layers in the encoder. The first hidden layer of the CFNN is a linear layer that maps the system states and control inputs to the scheduling parameters using the following relation:
%
%
\begin{equation}
c^{[1]}(i) = \upsilon (i) = W^{[1]}w(i)\quad \text{for } i\in \mathbb{N}_N
\label{eqn5_2_2}
\end{equation}
where $W^{[1]} \in \mathbb{R}^{l \times (n+n_u) }$ is the known weight matrix and $c^{[1]}$ is the output of the first hidden layer and the input to the map $\eta$, with $c^{[1]}(i)\in \mathbb{R}^{l} $.
The output of each of the other $n_e{-}1$ layers in the encoder is given as
\begin{equation}
    c^{[k]}(i) = g^{[k]}(W^{[k]}c^{[k-1]}(i)+b^{[k]}) \quad \text{for } i\in \mathbb{N}_N
\label{eqn5_2_3}
\end{equation}
where $k\in \mathbb{N}_{n_e}\backslash\{1\}$, $g^{[k]}$ is the nonlinear activation function, and $W^{[k]}$ and $b^{[k]}$ are the weight and bias of the $k^{th}$ hidden layer, respectively.
$c^{[n_e]}$ is the reduced parameter. The last hidden layer of the CFNN is the decoding layer, i.e., the map $\eta^\ast$ that reconstructs the original parameter from the reduced parameter using the following relation:
\begin{equation}c^{[n_e+1]}(i) = \hat{\upsilon}(i) = W^{[n_e+1]}c^{[n_e]}(i)+b^{[n_e+1]} \ \ \text{for } i\in \mathbb{N}_N.
\label{eqn5_2_4}
\end{equation}
%
%
We use linear activation in the decoder so that the reconstructed parameters have an affine dependence on the reduced parameters. This is done to ensure  the system matrix-valued functions obtained using \eqref{eqn5_1_6} have polynomial dependence on the reduced parameters. Since we use a linear decoder, the use of multiple decoding layers is unnecessary. The output of the CFNN is a concatenation of the input to the CFNN and the output of the last hidden layer. To learn the weights and biases, the following optimization problem is solved: 
\begin{equation}
\begin{aligned}
\min_{W^{[k]},\,b^{[k]}\, \text{for}\ k\in \mathbb{N}_{n_e+1}\backslash\{1\}} \quad & \frac{1}{N}\sum_{i=1}^{N}L_i(s_i, \hat{s}_i) + \bar{\sigma} \sum_{i=2}^{n_e+1}\left\lVert W^{[i]} \right\Vert_{F}^2\\
\end{aligned}
\label{eqn5_2_5}
\end{equation}
where $s_i^T=\begin{bmatrix}w(i)^T &\!\!\!\! \upsilon(i)^T \end{bmatrix}$ and
$\hat{s}_i^T=\begin{bmatrix}{w}(i)^T &\!\!\!\! \hat{\upsilon}(i)^T \end{bmatrix}$ are the target output and the output of the network, respectively. $L_i$ is a loss function that computes the
mean-squared
error between the state-output pairs of the original and the reduced LPV systems, and is defined as
$$L_i(s_i, \hat{s}_i) = \left\lVert \Big( M(\upsilon(i))-M(\hat{\upsilon}(i)) \Big)w(i) \right\Vert_{2}^2 .$$ Finally, $\bar{\sigma}$ is the regularization strength. The reduced LPV system can then be expressed as an LFT system with the scheduling parameters modeled as SLTV uncertainties.

\section{REDUCTION ERROR} \label{reductionerror}

\begin{figure}[tb]
\centering
\includegraphics[scale=0.3]{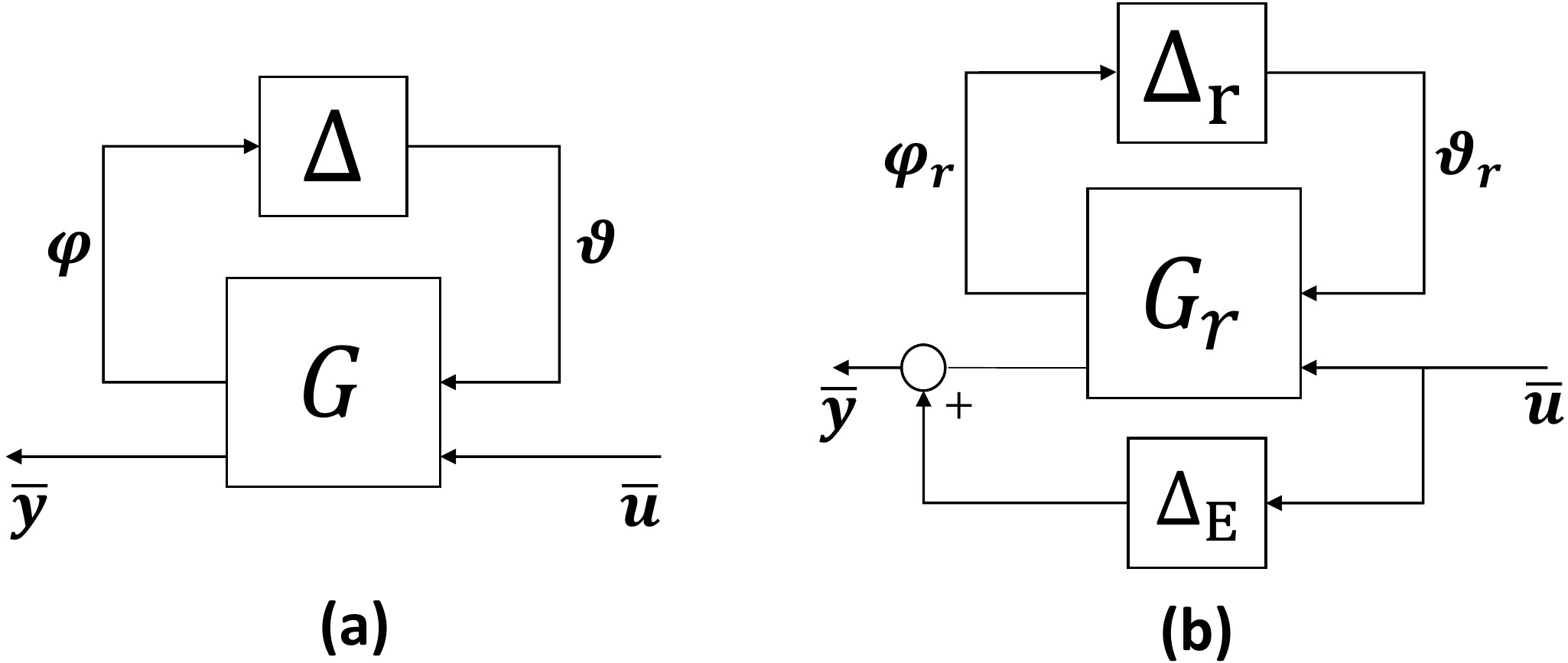}
\caption{a)~LFT representation of the LPV system; b)~Reduction error incorporated as dynamic uncertainty  $\Delta_E$.}
\label{fig_rLFT_wDLTI}
\end{figure}
%
%
%
%
By reducing the number of parameters and suitably modifying the parameter bounds based on the nonlinear map $\eta$ obtained from the CFNN, one might end up with an LFT with a smaller bound on its $\ell_2$-induced norm as obtained from IQC analysis. However, this LFT representation might not capture the nonlinear system dynamics as satisfactorily as the original LFT.
Herein, a dynamic uncertainty $\Delta_E$ is added to the LFT (see Fig.~\ref{fig_rLFT_wDLTI}b) to account for the  reduction error, thereby improving the model accuracy.
%
%
Assuming $\Delta_E$ is linear, to find the most appropriate norm bound
on $\Delta_E$,
one needs to compute the supremum of the $\ell_2$-induced norm of the error LFT system over all permissible $\Delta$. For each $\Delta$, the $\ell_2$-induced norm is finite if the error LFT system is stable, and for zero initial conditions, it is defined as
\begin{equation}
\begin{aligned}
    \lVert (G,\Delta) - (G_r,\Delta_r) \rVert_{\ell_2\rightarrow\ell_2} =  \sup_{\bar{u}\in \ell_2,\, \bar{u} \neq 0}  &  \frac{\lVert \delta \bar{y} \rVert_{\ell_2}}{\lVert \bar{u} \rVert_{\ell_2}}
\end{aligned}
\label{6_0_1}
\end{equation}
where $\delta \bar{y}$ is the error between the output histories of the original and the reduced LFT systems for the same input history $\bar{u}$.
%
%
The dynamic uncertainty can also account for the PNLSS modeling error if $(G,\Delta)$ is replaced with the nonlinear system \eqref{eqn3_0} and the norm used is the $\ell_2$-gain.

%
%
A simple approach to finding an approximate value of this bound
is to conduct multiple simulations with different input histories and simulation parameters, and then choose the maximum of the ratio of the error norm to the input norm over all the simulations. However, doing unguided simulations that span the envelope $\mathbb{H}$  is a cumbersome task, and unless a large number of simulations are carried out, the chances of obtaining a value close to the supremum in this way are slim.
%
%
To overcome this problem, we utilize falsification to perform guided simulations.
%
%

\subsection{Falsification}
In this work, we use falsification in a novel way to do guided simulations, where we attempt to maximize an objective. The falsification experiments are conducted using the S-TaLiRo toolbox \cite{staliro}. 
We define the system simulator as a black-box model that simulates the error system for a predefined amount of time. Following a similar definition to the $\ell_2$-induced norm, the initial conditions are set to zero for each simulation. The input history and other simulation parameters are judiciously sampled by S-TaLiRo from a user-defined search space to minimize the robustness metric. The robustness metric is the measure of how close the simulation results are to violating a specified property.
The simulator outputs the ratio  of  the  error  norm  to  the input norm at the end of each simulation run, and the property to be violated is
that the simulator output is less than $d$, where $d$ can be any real number.
By minimizing the robustness metric, we maximize the output of the simulator, and the maximum value obtained is selected as the dynamic uncertainty bound.

When using falsification for CFNN training data generation, the following changes are made: a) the black-box model simulates the nonlinear system; b) the output of the black-box model is a metric that quantifies the uniformity of the samples generated through simulation; and c) the initial conditions are non-zero and  sampled by S-TaLiRo for each simulation. This algorithm is further described in Section~\ref{casestudy}.

\subsection{Reduced Parameter System} \label{orps}
Reducing the number of parameters  will increase the norm bound on the dynamic uncertainty and change the size of the SLTV uncertainty block,  thus affecting the computational complexity of the analysis problem and  the conservativeness of the analysis results.
IQC-based analysis tools  can be used to analyze the uncertain LFT system; see \cite{Megretski1997} for an overview. The IQC-based analysis approach can handle a wide range of uncertainties, including static and dynamic, LTI and linear time-varying  perturbations, sector-bounded nonlinearities, and time delays. The systems of interest  are expressed as the interconnection $(G,\Delta_c)$ of a causal, stable, LTI nominal system $G$ and a perturbation operator $\Delta_c$ that belongs to a set $\bm{\Delta}_c$ of causal, bounded operators on $\ell_2$. Given an LFT $(G,\Delta_c)$ that maps the input $\bar{u}\in \mathcal D \subseteq \ell_2$ to the output $\bar{y} \in \ell_2$ for $\Delta_c\in\bm{\Delta}_c$, the uncertain system  $(G,\bm{\Delta}_c)$ is said to have a robust $\mathcal{D}$-to-$\ell_2$-gain performance level of $\gamma$ if it is robustly stable and  $\lVert (G,\Delta_c) \rVert_{\mathcal{D}\rightarrow\ell_2} \,{<}\, \gamma$ for all $\Delta_c \in \bm{\Delta}_c$.
In this work, we employ IQC analysis to compare the different uncertain LFT systems. Based on the $\gamma$-value obtained from IQC analysis and the size of
the LFT,
we select the appropriate  LFT representation of the nonlinear system over the envelope~$\mathbb{H}$.

\section{CASE STUDY} \label{casestudy}
\begin{figure}[tb]
\centering
\includegraphics[width=0.38\textwidth]{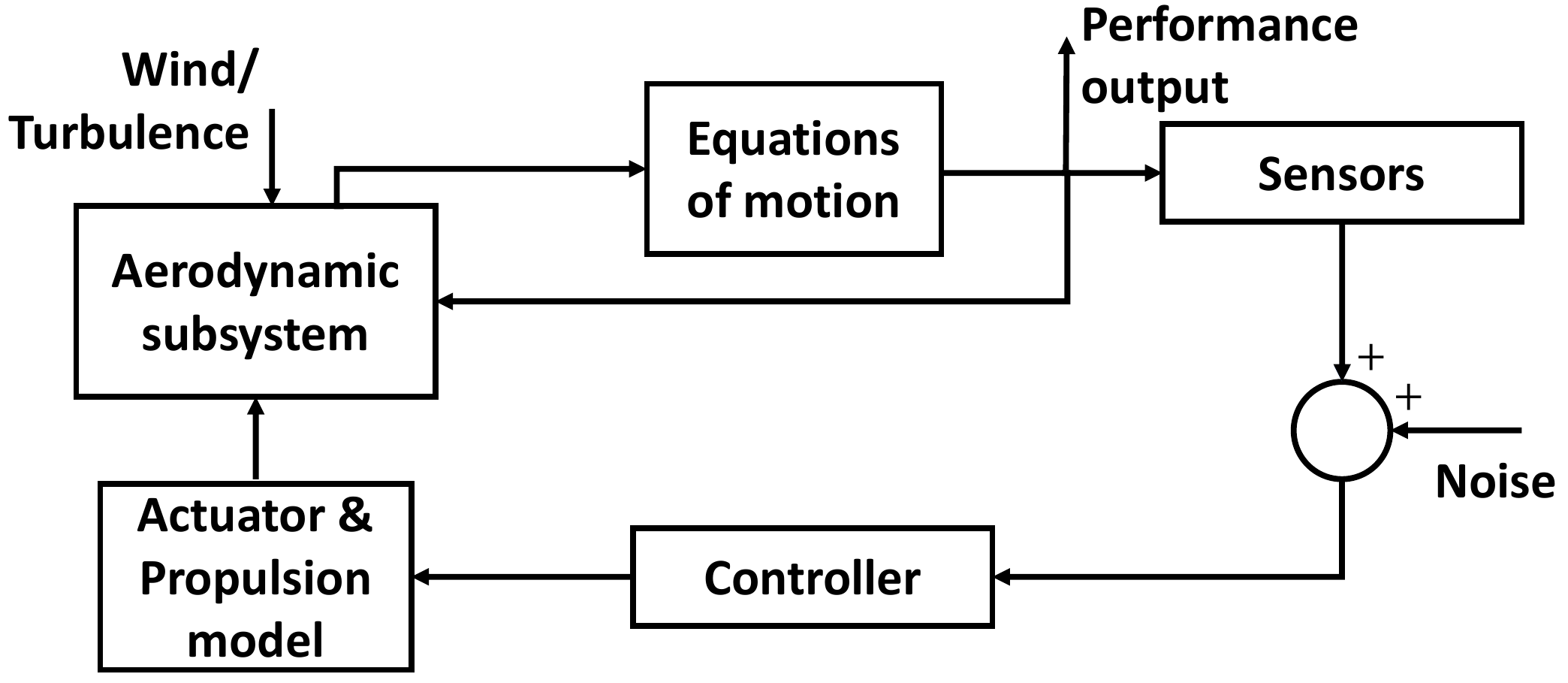}
\caption{The fixed-wing UAS.}
\label{fig_UASflowchart}
\end{figure}

We consider the problem of representing the nonlinear equations of motion (EOM) of a UAS with an uncertain LFT system as a case study. The uncertainties associated with the other subsystems are not considered here; see \cite{Palframan2017} for a detailed investigation on how to model these uncertainties.
All the subsystems of the UAS (Fig.~\ref{fig_UASflowchart}) except the controller are modeled as in \cite{Muniraj2017}.
The EOM of the UAS is expressed in the form of equation \eqref{eqn3_0} with $x = [\omega^T,\, {\bf v}^T,\, \lambda^T,\, p^T]^T$ and $u = [F^T,\,M^T]^T$. The state equation \eqref{eqn3_0_1} is nonlinear in  ${\omega}$, ${{\bf v}}$, and  ${\lambda}$. The output equation \eqref{eqn3_0_2} is linear since the output of the EOM subsystem is the same as the state.

{\it 1) LPV Representation:}
%
%
The EOM is linearized about a straight-and-level trim and then discretized. Samples of ${\omega}$, ${{\bf v}}$, and  ${\lambda}$ that are required to solve the regression problem in \eqref{eqn3_3_1} are sampled from $\mathbb{H}$ using the Halton method \cite{Perdegnana2016}. The optimization problem is solved using the \textsc{CVX} toolbox~\cite{Grant2014}.
%
%
%
%
To factorize $ {E}^T{\zeta}(\bar{x},\bar{u})$ as $\Delta A_d(\rho)\bar{x} $, a priority order is defined with elements of $\bf v$ on top. The maximum degree of these variables in the monomial function is one, and no monomial has more than one of these three variables. This results in a $6$-parameter LPV system with $\rho = [\bar{\omega}^T,\,\bar{\lambda}^T]^T$.
%
%
%

{\it 2) Parameter Reduction: }
%
%
To reduce the number of parameters, we implement the CFNN framework in Keras \cite{keras} with $n_e = 2$. A \texttt{tansig} activation function is chosen for the encoding layer. The network is trained using the Adam optimizer \cite{Adam2014} with a minibatch size of $128$ and a regularization strength of $10^{-6}$.
%
%
Once the CFNN is trained, the training loss $(T_L)$ and validation loss $(V_L)$ can be defined as the mean squared error between the state-output pairs of the original and the reduced LPV systems calculated using the training data and the validation data, respectively. The losses for different values of $m$ are provided in Table~\ref{tbl_trainingloss}.

The training and validation samples are generated through closed-loop simulations of the nonlinear system. The closed-loop system (Fig.~\ref{fig_UASflowchart}) takes as input the disturbance $d$ consisting of wind (steady component and turbulence) and measurement noise. The performance output of the closed-loop system is the output of the EOM subsystem, i.e., $\bar{x}$. An LTI $\mathcal{H}_\infty$ trajectory-tracking controller is designed following the procedure outlined in \cite{Gahinet1994}. Since the nonlinearity is due to the state variables, we only need to generate trajectories of $\bar{x}$, as the terms with input variables get canceled in the loss function. To better explore the state space, instead of just tracking the straight-and-level trajectory, we track a parameter-varying flight trajectory. The parameters that influence the reference trajectory are the flight path angle (FPA) and the radius of curvature (ROC). The ROC at each point is obtained from the desired heading angle $(\bar{\psi})$ history. This is done to ensure that the heading angle lies within $\mathbb{H}$. The measurement noise and the input to the turbulence model are sampled from a Gaussian white noise distribution.

We use falsification to sample the initial conditions, the steady wind, and the FPA and reference heading angle histories for each simulation with the objective of maximizing the uniformity of the samples within the envelope. One way of quantifying the uniformity of the samples is to compute the modified $L_2$-star discrepancy (ML$_2$) metric \cite{Perdegnana2016}; however, calculating this metric is computationally expensive. Instead, we maximize the spread of the samples from the operating point, which is quantified by $V_i$ defined as $\sum_{k=1}^{n}\sum_{j=1}^{s} \bar{x}_{kj}^2$ for a simulation run $i$, where $s$ is the total number of samples generated from the simulation run. Since we are only concerned with the envelope $\mathbb{H}$, we remove the samples that are outside $\mathbb{H}$ and sort the simulation runs with respect to $V_i$. Simulation runs with low $V_i$ values are not desired, as they contain samples that are clustered around the operating point. So, we define a lower bound $L_v$ for $V_i$ and reject simulation runs with $V_i<L_v$. We then create a simulation database and then select simulation runs from the database using a uniform gridding method and add samples until we see that any further addition of samples does not lead to an increase
%
%
in the ML$_2$ metric.
%
%

%
%
%
%

\begin{table}[t]
\renewcommand{\arraystretch}{1.3}
\centering
\begin{tabular}{|c|c|c|c|c|c|}
\hline
\bfseries m & \bfseries 5 & \bfseries 4 & \bfseries 3 & \bfseries 2 & \bfseries 1\\
\hline
$T_L\, (\times 10^{-8})$ & $6.19$ & $22.6$ & $58.0$& $153$& $213$\\
\hline
$V_L\, (\times 10^{-8})$ &  $7.01$ & $23.3$&$59.2$ & $157$& $241$\\
\hline
\end{tabular}
\caption{Network training and validation loss.}
\label{tbl_trainingloss}
\vspace{-2mm}
\end{table}

{\it 3) Reduction Error:}
%
%
The EOM subsystem is unstable, and so we define $\Delta_E$ to be a mapping from the disturbance input to the performance output of the closed-loop UAS. The uncertainty block $\Delta_E$ can be represented by a $12\times 13$ transfer matrix.
We employ falsification to compute the appropriate bound on
$\Delta_E$ as discussed in Section~\ref{reductionerror}. As in the previous subsection, we use parameter-varying trajectories to better explore the envelope $\mathbb{H}$, and falsification is used to sample the steady wind and the FPA and reference heading angle histories.
Table~\ref{tbl_DLTIbound} provides the bounds on $\Delta_E$. For the original $6$-parameter LPV system, the size of $\Delta(k)$ is $45\times 45$, and $\Delta_E$ is non-zero because of the PNLSS modeling error. We do not consider the $5$, $4$, and $3$ parameter LPV systems because  the size of  $\Delta_r(k)$  increases from $45\times 45$ to $134\times 134$, $99 \times 99$, and $68 \times 68$, respectively. For the LPV systems with 2 and 1 parameters, the size of $\Delta_r(k)$ decreases to $41\times 41$ and $18\times 18$, respectively. Note that, for the $1$-parameter LPV system, it might be possible to further reduce the size of $\Delta_r(k)$ without inducing any errors as shown in the example provided in~\cite{NSLPVcontrol}. To demonstrate the benefits of using an LFT system with the SLTV and dynamic uncertainties as opposed to an LFT system with just the dynamic
%
%
uncertainty to represent the nonlinear system, we also analyze the LTI system ($m=0$) with the dynamic uncertainty.
%
%

%
%
%
%

%
%
%
%

\begin{table}
\renewcommand{\arraystretch}{1.3}
\centering
\begin{tabular}{|c|c|c|c|c|c|c|}
\hline
\bfseries m & \bfseries 0 & \bfseries 1 & \bfseries  2 &    \bfseries 6 \\
\hline
$  \max{ \lVert \delta \bar{y} \rVert_{\ell_2} / \lVert \bar{u} \rVert_{\ell_2}} $
& 1.12 & 0.47 & 0.29  &  0.11 \\
\hline
\end{tabular}
\caption{Norm bounds for the dynamic uncertainty.}
\label{tbl_DLTIbound}
\vspace{-2mm}
\end{table}

{\it 4) IQC Analysis:}
The IQC analysis framework developed in \cite{Palframan2017} for robustness analysis of UAS is utilized here to estimate the upper bound on the $\mathcal{D}$-to-$\ell_2$-gain performance level of the map from the disturbance input to the performance output for all permissible uncertainties. To generate the LFT of the closed-loop UAS, we linearize the aerodynamic, sensor, actuator, and propulsion models about the straight-and-level trim condition.
In IQC analysis, each sensor noise channel is characterized using a banded white signal IQC
within a frequency range of $[-2\pi/3,\, 2\pi/3]$. The values used for the poles of the basis function pertaining to the signal IQC multiplier are $-0.9$ and $0.9$.
The scheduling parameters are characterized as rate-bounded SLTV uncertainties, and the SLTV parameter bounds and the parameter increment bounds are obtained from the parameter trajectories generated from the simulations mentioned before. $\Delta_E$ is characterized as a possibly nonlinear, norm-bounded uncertainty using the IQC multiplier provided in \cite{veenman2016}, and the basis length and pole location are chosen as $1$ and $-0.6$, respectively.


%

The following four systems are considered during analysis: the three uncertain LFT systems corresponding to $m=1$, $2$, and $6$, and the LFT system with just the dynamic uncertainty, and for each of these systems, two cases are studied, namely, the case where the dynamic uncertainty is considered and the case where it is not.
The resulting semidefinite programs are solved in a matter of minutes on a standard laptop using  MOSEK \cite{mosek}.
The $\gamma$-values for both cases are given in Table~\ref{tbl_IQCgamma}.
%
%
For the LFT system corresponding to $m=0$, there is about a $5$-fold increase in the value of $\gamma$ when the dynamic uncertainty is incorporated, which indicates that this LFT representation might be quite conservative.
If the dynamic uncertainty is not considered, then as we reduce the number of parameters, the $\gamma$-values obtained from IQC analysis also decrease, which is not surprising. However, if the dynamic uncertainty is incorporated into the analysis, then the resulting $\gamma$-values become
%
%
comparable for the different LFT systems with SLTV and dynamic uncertainties,
as evident from Table~\ref{tbl_IQCgamma}. Based on these observations, it seems that the LFT system with one parameter and a dynamic uncertainty constitutes a
relatively good representation of the nonlinear system over $\mathbb{H}$ in the sense that this LFT has the
smallest size among all considered LFTs with SLTV uncertainties and the associated analysis results~are~not~overly~conservative.

\section{CONCLUSION}
This work provides an approach to
derive an uncertain LFT system that adequately captures the behavior of a nonlinear system over some envelope.
We develop a CFNN-based framework for reducing the number of uncertainties in the LFT system. The CFNN framework enables one to directly minimize the errors in the state and output equations while reducing the number of uncertainties. To account for the reduction error and the error associated with PNLSS model estimation, we add a dynamic uncertainty to the LFT system. Guided simulations using falsification are used to obtain an upper bound on the dynamic uncertainty and to generate uniform data for parameter reduction. The benefits of the  approach are demonstrated in a case study involving the nonlinear equations of motion of a fixed-wing UAS.

%
%

\begin{table}[tb]
\renewcommand{\arraystretch}{1.3}
\centering
\begin{tabular}{|c|c|c|c|c|}
\hline
\bfseries m & \bfseries 0 &\bfseries 1 &\bfseries   2  &\bfseries  6 \\
\hline
$ \gamma~ (\gamma~ \text{w/o} ~\Delta_E)  $   & \!\!12.13 (2.53)\!\! & \!\!6.72 (2.78)\!\! & \!\!6.84 (4.31)\!\!  & \!\!7.19 (6.53)\!\! \\
\hline
\end{tabular}
\caption{Upper bounds on the $\mathcal{D}$-to-$\ell_2$-gain performance levels.}
\label{tbl_IQCgamma}
\vspace{-2mm}
\end{table}
%
%

\addtolength{\textheight}{-12cm}   








\end{document}